\documentclass[final,5p,times]{elsarticle}
\biboptions{numbers,sort&compress}
\usepackage{amssymb}
\usepackage{tikz}
\usepackage{graphicx}
\usepackage{array}
\usepackage{lipsum}
\usepackage{amsmath}
\usepackage{float}
\usepackage{amsthm}
\usepackage{amsfonts}
\usepackage{booktabs}
\usepackage{pgfplots}
\usepackage{pgfplotstable}
\usepackage{todonotes}

\usetikzlibrary{decorations.markings, arrows.meta}
\newtheorem{example}{Example}
\usepackage[colorlinks=true, allcolors=cyan]{hyperref}

\biboptions{square,comma}

\journal{Physics Letters B}

\begin{document}

\hypersetup{
    colorlinks=true,
    allcolors=cyan
}

\begin{frontmatter}



\title{BPS spectroscopy with reinforcement learning}


\author[a,b]{Federico Carta}
\author[a,c]{Asa Gauntlett}
\author[a,d]{Finley Griffin}
\author[a,e]{Yang-Hui He}

\affiliation[a]{
    organization={London Institute for Mathematical Sciences},
    addressline={Royal Institution},
    city={London},
    postcode={W1S 4BS},
    country={UK}
}

\affiliation[b]{
    organization={Physics Department, King’s College London},
    addressline={Strand},
    city={London},
    postcode={WC2R 2LS},
    country={UK}
}

\affiliation[c]{
    organization={Department of Computer Science, University College London},
    addressline={Gower Street},
    city={London},
    postcode={WC1E 6BT},
    country={UK}
}

\affiliation[d]{
    organization={Hertford College, University of Oxford},
    addressline={Catte Street},
    city={Oxford},
    postcode={OX1 3BW},
    country={UK}
}

\affiliation[e]{
    organization={Merton College, University of Oxford},
    addressline={},
    city={Oxford},
    postcode={OX1 4JD},
    country={UK}
}

\begin{abstract}
We apply reinforcement learning (RL) to establish whether at a given position in the Coulomb branch of the moduli space of a 4d $\mathcal{N} = 2$ quantum field theory (QFT) the BPS spectrum is finite. If it is, we furthermore determine the full BPS spectrum at such point in moduli space.  We demonstrate that using a RL model one can efficiently determine the suitable sequence of quiver mutations of the BPS quiver that will generate the full BPS spectrum. We analyse the performance of the RL model on random BPS quivers and show that it converges to a solution various orders of magnitude faster than a systematic brute-force scan. As a result, we show that our algorithm can be used to identify all minimal chambers of a given $\mathcal{N}=2$ QFT, a task previously intractable with computer scanning. As an example, we recover all minimal chambers of the $\text{SU}(2)$ $N_f = 4$ gauge theory, and discover new minimal chambers for theories that can be realized by IIB geometric engineering.
\end{abstract}




\end{frontmatter}



\def\cN{\mathcal{N}}
\def\cZ{\mathcal{Z}}
\def\IC{\mathcal{C}}

\section{Introduction and summary}
\label{introduction and summary}
The Seiberg-Witten solution of $4$d $\mathcal{N}=2$ quantum field theories (QFT)s determines the low energy dynamics on the Coulomb Branch  of the moduli space of vacua \cite{Seiberg:1994rs,Seiberg:1994v2}. One crucial feature of the QFT which is not implied by the Seiberg-Witten solution is the spectrum of BPS states at a generic point of the Coulomb Branch, and how such spectrum is modified by moving in moduli space.
\par
Determining the BPS spectrum of theories with $\cN=2$ supersymmetry is non-trivial and yet seemingly tractable in numerous examples. A subset of $4$d $\mathcal{N}=2$ admits a BPS quiver \cite{Cecotti:2011rv,Alim:2011ae}, from which the BPS spectrum can be determined via various techniques. The problem becomes even simpler for a class of four-dimensional, $\cN=2$ supersymmetric QFTs which are dubbed {\em complete} in \cite{Cecotti:2011rv,Alim:2011ae}. Such $\cN=2$ models are defined by the rather stringent property that even as one varies all parameters such as moduli, couplings and bare masses, the number of hypermultiplets which constitute a basis of the BPS spectrum remains to be equal to the rank of the lattice of electromagnetic and flavour charges $\Gamma$.
\par
An example of a complete QFT is that obtained from two coincident M5-branes on a punctured Riemann surface, which gives us the theory constructed by Gaiotto \cite{Gaiotto:2009we}, composed of a product of $\text{SU}(2)$ gauge groups with tri-fundamental matter. Various other theories of the class $\mathcal{S}$ construction fall into the set of complete QFTs. In addition to these, Cecotti and Vafa \cite{Cecotti:2011rv} identified 11 exceptional cases, related to Dynkin diagrams for affine and elliptic exceptional groups as well as the Derksen-Owen quivers from \cite{DerksenOwen}.

Now, as we enter the age of AI \cite{kissinger2021age}, it was inevitable that machine learning (ML) methods should enter theoretical physics, especially in high-energy theory, and pure mathematics: this was introduced in \cite{He:2017aed,Krefl:2017yox,Carifio:2017bov,Ruehle:2017mzq} (for pedagogical introductions, see \cite{He:2018jtw,Ruehle:2020jrk} and for recent reviews on this ``AI-guided theoretical discovery'', see \cite{gukov2024rigor,He:2024gnk}).

The use of ML is by now well-established in string theory and related sub-fields of mathematical physics such as conformal fieldtheory \cite{Chen:2020dxg,Kantor:RL}, $\mathcal{N}=1$ quiver QFTs \cite{Bao:seibergML}, cluster algebras \cite{Cheung:2022itk,he2024machines}, and BPS indices \cite{Gukov:2024opc}.
It is therefore natural and expedient to see whether AI, and in particular ML methodologies, can help with our present problem of understanding BPS spectra and the classification of $\cN=2$ gauge theories.

This above question we address in the present paper, whose main results we summarize here:
\begin{enumerate}
    \item As a main result, we develop a reinforcement learning (RL) algorithm which efficiently determines a sequence of quiver mutations that identifies the full BPS spectrum of a complete QFT, in any finite chamber of the moduli space. The code is available on \href{https://github.com/fingriffin}{GitHub}.
    \item We apply this algorithm to the $\text{SU}(2)\; N_f=4$ theory. We determine the number of inequivalent chambers in the Coulomb Branch for which the BPS spectrum is finite and minimal. To demonstrate the efficiency of our algorithm, we determine the BPS spectrum in all such chambers. These are also included in the \href{https://github.com/fingriffin}{GitHub}.
    \item As another example, we identify minimal chambers and corresponding BPS spectra for theories whose BPS quiver is given by the elliptic $E$-type Dynkin diagrams, $\hat{\hat{E}}_n$ with $n = 6,7,8$ as well as the theories identified by the Derksen-Owen quivers $X_6,X_7$ \cite{DerksenOwen}.  
\end{enumerate}

Our results represent a tremendous improvement to the efficiency and scalability of conventional methods for determining finite chambers and BPS spectra of $\mathcal{N}=2$ QFTs. Such a tool should aid in the current problem of the classification of $\mathcal{N}=2$ models and, furthermore, the study of their non-perturbative properties. 
For example, the mutation sequences identified in each case may be used as input data to the Cordova-Shao algorithm \cite{Cordvoa:2016} to determine their Schur index. 

It is also interesting to try and explain our conjecture for the number of minimal chambers, see Fig. \ref{fig:su2table}, for the $\text{SU(2)}\;N_f=4$ model from study of the underlying physics or pure mathematics. Finally, our computations may also be extended to compute non-minimal, or even maximal, finite chambers of $\mathcal{N}=2$ QFTs.

\section{Background}
We begin with a brief overview on how the BPS spectrum of complete theories can be derived from the BPS quiver.
\subsection{BPS quivers of $4D\;\mathcal{N}=2$ quantum field theories}

Let us consider a 4d $\cN=2$ QFT with a gauge group of rank $r$ and flavor symmetry algebra of rank $f$. The moduli space of supersymmetric vacua is split into a Coulomb branch $\mathcal{M}_C$ and a Higgs branch $\mathcal{M}_H$. On a generic vacua $u \in \mathcal{M}_C$ the gauge group is broken to $U(1)^r$. The lattice of electric, magnetic and flavour charges $\Gamma$ is of dimension $2r+f$ and is equipped with a linear function 
\begin{equation}
Z_u:\Gamma\rightarrow \mathbb{C}\;,
\end{equation}
called the central charge of the theory. 

By convention, the complex plane, where $Z_u$ takes values, is (arbitrarily) split into two half-planes $\mathbb{C}_{\pm}$. States with charges $\gamma\in\Gamma$ such that $Z_u(\gamma)\in \mathbb{C}_{+}$ (respectively, $Z_u(\gamma)\in \mathbb{C}_{-}$) are labelled as particles (respectively, antiparticles) of the theory. Furthermore, the charge lattice is equipped with an antisymmetric inner product 
\begin{equation}
\circ : \Gamma \times \Gamma \to \mathbb{Z}\;.
\end{equation}
At a point $u$ on the Coulomb Branch, the $\mathcal{N}=2$ superalgebra imposes the mass constraint on particles with charge $\gamma\in \Gamma$ 
\begin{equation}
M\geq |Z_u(\gamma)|\;.
\end{equation}
Particles which saturate this inequality are called {\it BPS particles}. For every point $u\in \mathcal{M}_C$, a relevant and important problem is finding the full spectrum of BPS particles, including as well their spin and multiplicity. For a subset of the $4d$ $\cN=2$ theories, such a question can be answered with the auxiliary tool of a BPS quiver. 

The BPS quiver is a $1d$ unitary quiver gauge theory living on the worldline of BPS particles that describes their dynamics. It can be defined as follows.
Let $\{\gamma_i\}$ be the charges of the hypermultiplets which form a basis of the BPS spectrum. Then:
\begin{itemize}
    \item Each element $\gamma_i$ in the basis corresponds one node of the quiver.
    \item For each pair of charges in the basis, compute the electric-magnetic inner product $\gamma_i \circ \gamma_j$ . If $\gamma_i \circ \gamma_j > 0$, the nodes $\gamma_i$ and $\gamma_j$ are connected with $\gamma_i \circ \gamma_j$ arrows  \footnote{If $\gamma_i \circ \gamma_j < 0$, then we have no arrows. Note that this convention is slightly different from that in the quiver gauge theory literature where the antisymmetrized adjacency matrix is used so that a negative intersection means an arrow going the other way.}, each pointing from node $j$ to node $i$. 
\end{itemize} 

From the BPS quiver alone it is possible to determine the BPS spectrum at any point in the moduli space. One way of doing so consists in determining the stability conditions of various quiver representations, each one corresponding to a particle of the QFT. We will not include a detailed discussion of this method, and instead refer the reader to \cite{Alim:2011kw}. A much simpler method involves a computation of a sequence of quiver mutations, and is dubbed the \textit{mutation method} \cite{Alim:2011kw}. We will review this method in the next subsection.

\subsection{Quiver mutation and finite chambers}

If at a point $u \in \mathcal{M}_C$ the BPS spectrum is finite, the spectrum can be fully determined by the so-called mutation method \cite{Alim:2011kw}. A \emph{quiver mutation} is an operation \footnote{In the mathematics literature, this was coined by Fomin-Zelevinsky in the context of cluster algebras \cite{fomin2002cluster}. Interestingly, in physics, this was realized independently around the same time \cite{Feng:2000mi} as Seiberg duality or toric duality. It was only years later that at a Oberwolfach workshop that the two communities realized that they were talking about the same thing.} that is applied at a node of a BPS quiver $Q$. After the mutation, a new BPS quiver $\tilde{Q}$ is produced. The new quiver $\tilde{Q}$ will have the same number of nodes, although now labeled with different charges $\tilde{\gamma}_i$, and a different set of arrows. 

In order to find the full BPS spectrum one needs to apply a sequence of quiver mutations (generically acting on different nodes) such that at the end of the process, the final quiver produced will have all charges flipped compared to the original quiver and all arrows inverted. We call such end-point of the mutation method the \emph{antiparticle quiver}. Keeping track of all the charges that appear at all nodes in every intermediate steps is equivalent to determine the full BPS spectrum. We will recall below the rules defining the mutation operation. 
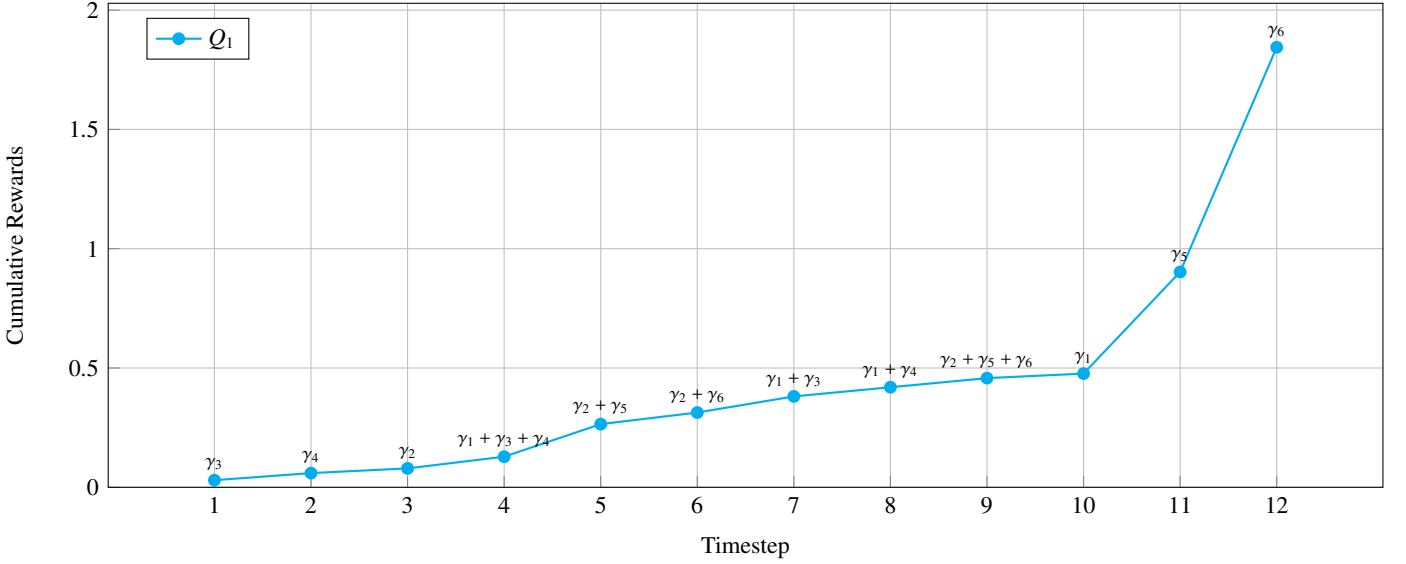
\begin{figure*}[h]
    \centering
    \begin{tikzpicture}
\begin{axis}[
    title={Cumulative Reward over Time; $\text{SU}(2)$ $N_f = 4$},
    xlabel={Timestep},
    ylabel={Cumulative Rewards},
    ymin=0,
    legend pos=north west,
    grid=both,
    label style={font=\small},
    tick label style={font=\small},
    title style={font=\small},
    legend style={font=\small},
    xtick pos=left,
    ytick pos=left,
    xtick=data,
    width=\textwidth,
    height = 8cm,
    nodes near coords={
        \ifnum\coordindex=0 $\gamma_3$\fi
        \ifnum\coordindex=1 $\gamma_4$\fi
        \ifnum\coordindex=2 $\gamma_2$\fi
        \ifnum\coordindex=3 $\gamma_1+\gamma_3+\gamma_4$\fi
        \ifnum\coordindex=4 $\gamma_2+\gamma_5$\fi
        \ifnum\coordindex=5 $\gamma_2+\gamma_6$\fi
        \ifnum\coordindex=6 $\gamma_1+\gamma_3$\fi
        \ifnum\coordindex=7 $\gamma_1+\gamma_4$\fi
        \ifnum\coordindex=8 $\gamma_2+\gamma_5+\gamma_6$\fi
        \ifnum\coordindex=9 $\gamma_1$\fi
        \ifnum\coordindex=10 $\gamma_5$\fi
        \ifnum\coordindex=11 $\gamma_6$\fi
    },  
    every node near coord/.append style={font=\scriptsize, color=black}  
]
    \addplot[color = cyan, thick, mark=*] table [x=timestep, y expr=\thisrow{actual_reward}/100, col sep=comma] {data/fig1data.csv};
    \addlegendentry{$Q_1$}
\end{axis}

\end{tikzpicture}
    \caption{Cumulative reward ($\lambda = 0.1$) against mutation for a successful training episode on the BPS quiver $Q_1$ of the $\text{\text{SU(2)}} \; N_f = 4$ SCFT, from Fig. \ref{fig:su2table}, identifying a finite chamber containing $12$ states.}
    \label{fig:cumreward}
\end{figure*}
To define the mutation, let us suppose that we mutate at the node labeled by $1$, and whose corresponding charge is $\gamma_1$. Let us call $Q$ the original quiver, and $\tilde{Q}$ the quiver after the mutation. The procedure to derive $\tilde{Q}$ from $Q$ is then the following:
\begin{enumerate}
    \item The nodes of the mutated quiver $\tilde{Q}$ are the same of the nodes of the quiver $Q$. Also, for every arrow in the quiver $Q$, write the correspondent arrow in the quiver $\tilde{Q}$.
    \item For each length two path of arrows passing through node 1 in $Q$, draw a new arrow in $\tilde{Q}$ connecting the initial and final node of the two steps path.
    \item Invert all arrows in $\tilde{Q}$ which end on node 1.
    \item If between any two nodes $i$, $j$ of the quiver $\tilde{Q}$ there is both an arrow $i\to j$ and one $j\to i$, delete both of them. Repeat this procedure until no such couple of lines with opposite orientation exist anymore, for all couple of nodes.
    \item  The new basis is given by
\begin{eqnarray}\label{mutbasis}
\widetilde{\gamma}_1 & = & -\gamma_{1}\\
\widetilde{\gamma}_j  & = & 
  \begin{cases}
   \gamma_{j}+( \gamma_{j} \circ   \gamma_{1})\gamma_{1} & \text{if }  \gamma_{j} \circ   \gamma_{1} >0 \\
   \gamma_{j}      & \text{if }   \gamma_{j} \circ \gamma_{1} \leq0.
  \end{cases}
\end{eqnarray}
\end{enumerate}
There are further rules to derive the superpotential of the mutated quiver $\tilde{Q}$, but we will not make use of those in this paper.

We have reviewed here the mutation mechanism to determine the BPS spectrum from the BPS quiver. Such a method is of course much more efficient than the method of studying representation theory of the BPS quiver. However, crucially, for more complicated quivers guessing by eye the right sequence of mutations is very challenging, and also an extremely inefficient problem to be treated by a computer code which uses a brute-force algorithm. We will solve both of these issues by using Reinforced Machine Learning (RL) to compute the BPS spectrum in finite chambers. In the following sections, we outline the RL algorithm we use.

\begin{figure*}
    \centering
    \begin{tikzpicture}
\begin{axis}[
    title={Algorithm Performance on Random Cyclic BPS Quivers},
    xlabel={Number of Nodes},
    ylabel={Average Convergence Rate to Finite Chamber},
    ymode=log,
    ymin=0,
    minor y tick num=8, 
    legend pos=north west,
    grid=both,
    label style={font=\small},
    tick label style={font=\small},
    title style={font=\small},
    legend style={font=\small},
    xtick pos=left,
    ytick pos=left,
    ytick={1, 10, 100, 1000, 10000, 100000, 1000000, 10000000, 100000000, 1000000000, 10000000000, 100000000000},
    xtick=data,
    width=\textwidth,
    height = 8cm
]
    \addplot[color = cyan, thick, mark=*] table [x=num_nodes, y=avg_PPO, col sep=comma] {data/fig2data.csv};
    \addlegendentry{RL}

    \addplot[color = red, thick, mark=*] table [x=num_nodes, y=avg_RW, col sep=comma] {data/fig2data.csv};
    \addlegendentry{SW}

    \addplot[color = red, dashed, thick] table [x=num_nodes, y=avg_RW_proj, col sep=comma] {data/fig2data.csv};
    \addlegendentry{SW (projected)}
\end{axis}

\end{tikzpicture}
    \caption{Plot of of the average convergence rate to a finite chamber against number of nodes of a random BPS quiver for a smart walk (SW) and reinforcement learning model (RL).}
    \label{fig:convergence}
\end{figure*}
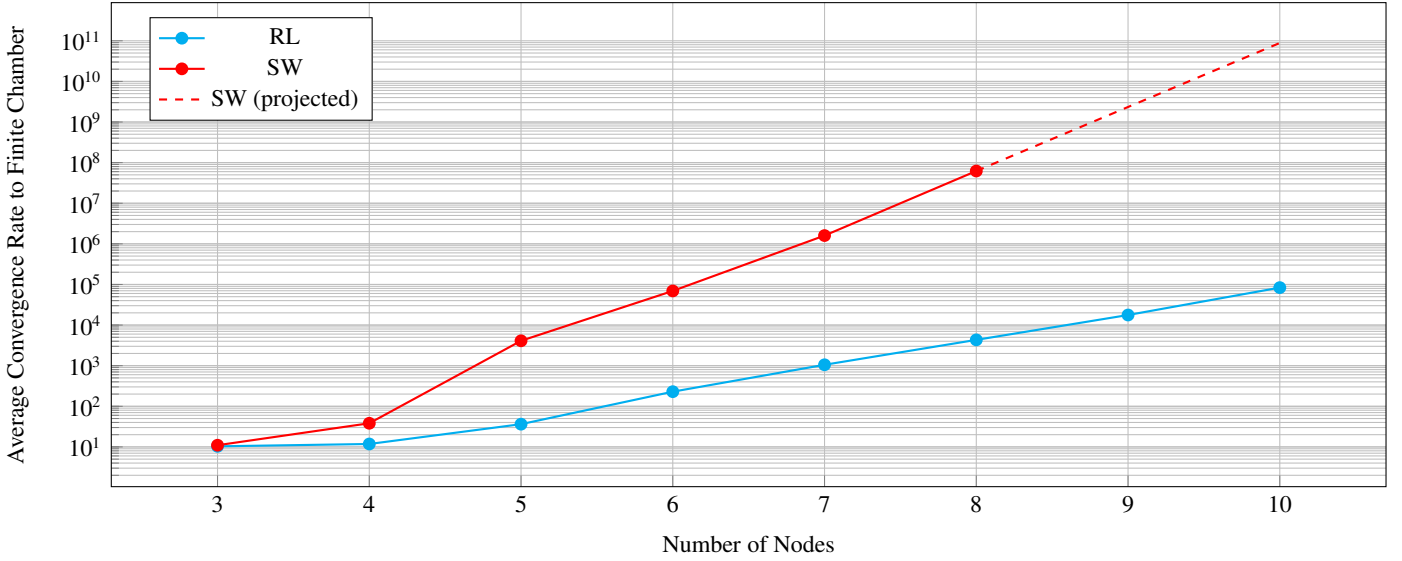

\section{Methods}
In this work, we use a Proximal Policy Optimisation (PPO) algorithm to address the problem of finding finite chambers of $\mathcal{N} = 2$ gauge theories. PPO is a model-free reinforcement learning (RL) involving an optimal policy $\pi^*: \mathcal{S}\rightarrow \mathcal{A}$ that maps any state $s \in \mathcal{S}$ to an action $a \in\mathcal{A}$. Given a state $s_t$ at discrete timestep $t$, the PPO algorithm selects an action $a_t$, computes a reward and moves to a subsequent state $s_{t+1}$. The policy $\pi$ is updated based on the clipped objective function and is represented by a deep neural network (NN). The NN we used for the optimisation procedure consists of three hidden layers of 252, 504 and 252 neurons with ReLU activation. The NN was trained using the Adam optimiser and the learning rate $\alpha$ and discount factor $\gamma$ were set to $0.5$ and $0.995$, respectively. Further details of RL and PPO can be found in \ref{RL}.
\par
In our investigation, training is initialized with an $n$-node BPS quiver $s_0 \in \mathcal {S}$, with $\mathcal{S}$ the space of BPS quivers equivalent to $s_0$ by quiver mutation. A state is represented by an $n\times 2n$ matrix $(M\;|\; \Gamma)$, where $M$ is the adjacency matrix of the underlying digraph, and $\Gamma$, the ‘gamma matrix’, encodes quiver labels as linear combinations of BPS hypermultiplets $\{\gamma_i\}$, i.e.
\begin{equation}
\text{node label }i = \sum_j\Gamma_{ij}\gamma_j\;;\;\Gamma_{ij} \in \mathbb{Z}\;. 
\end{equation}
\begin{example}\label{ex: 1}
Let $s_0$ be the BPS quiver $Q_1$ for the  $\text{\text{SU(2)}} \; N_f =4$ gauge theory \cite{Alim:2011kw} from Fig. \ref{fig:su2table}. Training is initialized with the data structure
\begin{equation}s_0=
\left(\begin{array}{cccccc|cccccc}
   0 & 2 & 0 & 0 & 0 & 0 & 1 & 0 & 0 & 0 & 0 & 0 \\
   0 & 0 & 1 & 1 & 1 & 1 & 0 & 1 & 0 & 0 & 0 & 0 \\
   1 & 0 & 0 & 0 & 0 & 0 & 0 & 0 & 1 & 0 & 0 & 0 \\
   1 & 0 & 0 & 0 & 0 & 0 & 0 & 0 & 0 & 1 & 0 & 0 \\
   1 & 0 & 0 & 0 & 0 & 0 & 0 & 0 & 0 & 0 & 1 & 0 \\
   1 & 0 & 0 & 0 & 0 & 0 & 0 & 0 & 0 & 0 & 1 & 1
\end{array}\right)\;.
\end{equation}
\end{example}
\noindent
The gamma matrix of the initial quiver $s_0$ is always set to the identity matrix, for example in Example \ref{ex: 1}, as the $i$-th node label is simply the hypermultiplet $\gamma_i$ alone. In general, intermediate quivers will have off-diagonal elements in $\Gamma$, representing bound states of $\{\gamma_i\}$. We also note the adjacency matrix $M$ adopts the standard convention, where an arrow from nodes $i \rightarrow j$ is represented by $M_{ij}=+1$, $M_{ji}=0$.

We define the action space $\mathcal{A}$ as all possible nodes that can be mutated on at a given timestep. Illegal nodes are those with labels that contain negative coefficients of the hypermultiplets, or that have already been mutated on in the action history. For this reason, the action space is dynamic, which makes the use of a PPO algorithm especially suitable\footnote{Our experiments showed that Deep-Q Learning proves much less stable with dynamic action masking in this context.}, as it demonstrates notable stability with action masking compared to other RL algorithms \cite{TangPPO}. 
\par
Finally, we define a reward function $R(s)$ that provides feedback for beneficial actions and a terminal reward ($+1$) when the antiparticle quiver is reached. A fitness function $f : \mathcal{S} \rightarrow [0,1]$ quantifies the proximity of a given intermediate state $s_t$ to the antiparticle quiver. We define two distinct fitness functions, $f_M$ and $f_\Gamma$, which separately evaluate this proximity in terms of the graph structure and node labels, respectively. The reward function is then
\begin{equation}
R(s) = 1-\lambda f_M(s)-(1-\lambda)f_\Gamma(s)\;,
\end{equation}
where $\lambda \in [0,1]$ is a tunable parameter representing the relative weighting of structural and label similarities in the reward logic. 
\begin{example}\label{ex:2}
Fig. \ref{fig:cumreward} shows the cumulative reward for a successful episode on the BPS quiver of the $\text{SU}(2) \; N_f = 4$ SCFT from Example \ref{ex: 1}, identifying the $12$-state chamber
\begin{equation}
\begin{array}{c}
\gamma_3,\gamma_4,\gamma_2,\gamma_1+\gamma_3+\gamma_4,\gamma_2+\gamma_5,\gamma_2+\gamma_6,\\\gamma_1+\gamma_3,\gamma_1+\gamma_4,\gamma_2+\gamma_5+\gamma_6,\gamma_1,\gamma_5,\gamma_6
\end{array}
\end{equation}
in agreement with the result from \cite{Alim:2011kw}. 
\end{example}

\section{Results}
To demonstrate the capability of our RL model to find a finite chamber for a generic theory, we established a dataset of cyclic digraphs, ranging from $3-10$ nodes, from which random BPS quivers are drawn. We then trained the RL algorithm on a random selection of these quivers, and recorded the convergence rate of the procedure as the number of time-steps to identify a finite chamber for a given quiver. A negative penalty is applied to episodes that exceed a maximum length or label complexity to encourage further exploration by the agent. As a baseline for performance evaluation, we use a smart walker (SW) algorithm that combines a random walk search with an exhaustive scan. This algorithm is equivalent to the RL algorithm with learning disabled. A log plot of the average convergence rate against number of nodes is shown in Fig.~\ref{fig:convergence}.

Some BPS quivers do not have finite chambers, which are identified by allowing the SW algorithm to run until it has exhausted all possible mutation sequences. Of the quivers with finite chambers, the RL algorithm achieves a $100\%$ accuracy and, moreover, is orders of magnitude more efficient than the SW algorithm. These results mean that the RL algorithm can efficiently identify finite chambers even for large BPS structures, where traditional exhaustive scans are intractable as the number of possible sequences scales as $\sim\text{(number of nodes)}^\text{length of sequence}$ \cite{Alim:2011kw}. 

For this reason, Fig.~\ref{fig:convergence} includes projections for the SW algorithm on $9$ and $10$ node quivers at $\mathcal{O}(10^9)$ and $\mathcal{O}(10^{10})$ time-steps, respectively. This would correspond to a computing time of $\sim 10$ days and $\sim 1$ year, respectively, on a home machine. We report an average of $\mathcal{O}(10^4)$ and $\mathcal{O}(10^5)$ time-steps with the RL algorithm, or a computing time of $\sim 5$ minutes and $\sim 20$ minutes, for $9$ and $10$ node quivers respectively.

Furthermore, the RL algorithm is inclined to identify minimal chambers of the theory due to the discount factor $\gamma$  in the $Q$-function, see \ref{RL}. For instance, the $12$-state chamber identified in Example \ref{ex:2} is indeed a minimal chamber of the theory. This is not a feature of the SW model, where the length of the chamber is needed a priori, and there is no guarantee that any finite chamber identified is a minimal one.

\subsection{Counting minimal chambers}
As discussed in \cite{XieBPSquivers}, it is interesting to find all the minimal chambers for all the BPS quivers of a given theory. To further showcase the capabilities of our model, we offer the original computation of all minimal chambers of the $\text{SU(2)} \; N_f = 4$ gauge theory. 

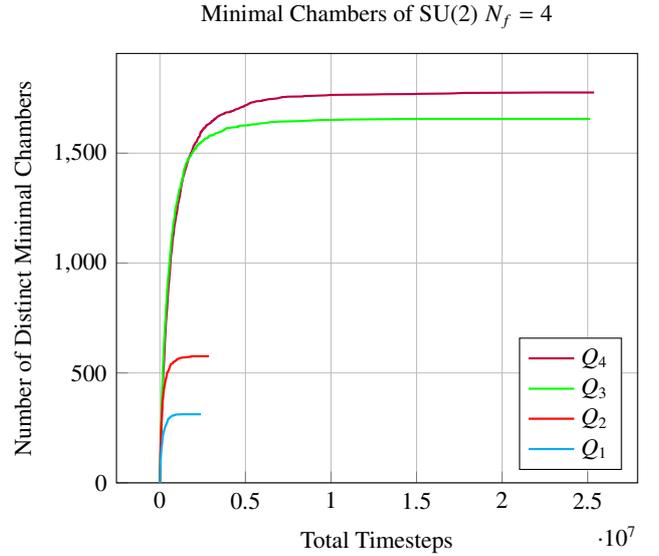
\begin{figure}[H]
    \centering
    \begin{tikzpicture}
\begin{axis}[
    title={Minimal Chambers of $\mathrm{SU}(2)\;N_f=4$},
    xlabel={Total Timesteps},
    ylabel={Number of Distinct Minimal Chambers},
    ymin=0,
    legend pos=south east,
    grid=both,
    label style={font=\small},
    tick label style={font=\small},
    title style={font=\small},
    legend style={font=\small},
    xtick pos=left,
    ytick pos=left,
]
    \addplot[color = purple, thick] table [x=Timestep, y=Number of Finite Chambers, col sep=comma] {data/fig4data_4.csv};
    \addlegendentry{$Q_4$}
    \addplot[color = green, thick] table [x=Timestep, y=Number of Finite Chambers, col sep=comma] {data/fig4data_3.csv};
    \addlegendentry{$Q_3$}
    \addplot[color = red, thick] table [x=Timestep, y=Number of Finite Chambers, col sep=comma] {data/fig4data_2.csv};
    \addlegendentry{$Q_2$}
    \addplot[color = cyan, thick] table [x=Timestep, y=Number of Finite Chambers, col sep=comma] {data/fig4data_1.csv};
    \addlegendentry{$Q_1$}

\end{axis}
\end{tikzpicture}
    \caption{Plot of the total number of distinct minimal chambers of $\text{\text{SU(2)}}\; N_f = 4$ found against the number of steps taken for a random walk search using the RL model with PPO.}
    \label{fig:all chambers}
\end{figure}

Minimal chambers for this theory have been identified many times 
\cite{Alim:2011kw,Alim:2011ae,XieBPSquivers,BrustleGreenSequences}, but the scale of an exhaustive computational scan means that a complete census is absent in the literature. For the $\text{SU(2)}\;N_f = 4$ theory, the Coulomb branch is spanned by four distinct BPS quivers, see Fig.~\ref{fig:su2table}, corresponding to the four possible triangulations of the fourth-punctured sphere \cite{Alim:2011kw,Cecotti:2011rv}. These can be verified using Keller's mutation Java applet \cite{KellerMutationApp}, as well as using our code.
\begin{figure}[H]
    \centering
    \begin{tabular}{|c|c|}
\hline
BPS Quiver & Minimal Chambers \\
\hline
\rule{0pt}{1.5cm}
$Q_1:$ \begin{tikzpicture}[>=stealth, node distance=2cm, baseline={(current bounding box.center)}]

\node[circle, draw=black, inner sep=4pt] (gamma3) at (-0.75*0.75, -0.5*0.75) {};
\node[circle, draw=black, inner sep=4pt] (gamma4) at (2.75*0.75, -0.5*0.75) {};
\node[circle, draw=black, inner sep=4pt] (gamma2) at (2.5*0.75, -1.73*0.75) {};
\node[circle, draw=black, inner sep=4pt] (gamma6) at (2.75*0.75, -2.96*0.75) {};
\node[circle, draw=black, inner sep=4pt] (gamma5) at (-0.75*0.75, -2.96*0.75) {};
\node[circle, draw=black, inner sep=4pt] (gamma1) at (-0.5*0.75, -1.73*0.75) {};

\node at (gamma3) {\footnotesize$\gamma_3$};
\node at (gamma4) {\footnotesize$\gamma_4$};
\node at (gamma2) {\footnotesize$\gamma_2$};
\node at (gamma6) {\footnotesize$\gamma_6$};
\node at (gamma5) {\footnotesize$\gamma_5$};
\node at (gamma1) {\footnotesize$\gamma_1$};

\draw[->, >=latex, transform canvas={yshift=1pt}] (gamma1) to (gamma2);
\draw[->, >=latex, transform canvas={yshift=-1pt}] (gamma1) to (gamma2);
\draw[->, >=latex] (gamma2) to (gamma3);
\draw[->, >=latex] (gamma2) to (gamma4);
\draw[->, >=latex] (gamma2) to (gamma5);
\draw[->, >=latex] (gamma2) to (gamma6);
\draw[->, >=latex] (gamma3) to (gamma1);
\draw[->, >=latex] (gamma4) to (gamma1);
\draw[->, >=latex] (gamma5) to (gamma1);
\draw[->, >=latex] (gamma6) to (gamma1);
\end{tikzpicture}& $312\color{lightgray}\times24\color{black}$
\\
\rule{0pt}{1.5cm}
$Q_2:$\begin{tikzpicture}[>=stealth, node distance=2cm, baseline={(current bounding box.center)}]

\node[circle, draw=black, inner sep=4pt] (gamma1) at (1.5*0.9+0.749*0.9, 0*0.9+1.06*0.9) {};
\node[circle, draw=black, inner sep=4pt] (gamma2) at (1.5*0.9, 0*0.9+2*1.06*0.9) {};
\node[circle, draw=black, inner sep=4pt] (gamma3) at (0*0.9,  0*0.9+2*1.06*0.9) {};
\node[circle, draw=black, inner sep=4pt] (gamma4) at (0*0.9-0.749*0.9, 0*0.9+1.06*0.9) {};
\node[circle, draw=black, inner sep=4pt] (gamma5) at (0*0.9, 0*0.9) {};
\node[circle, draw=black, inner sep=4pt] (gamma6) at (1.5*0.9, 0*0.9) {};

\node at (gamma1) {\footnotesize$\gamma_1$};
\node at (gamma2) {\footnotesize$\gamma_2$};
\node at (gamma3) {\footnotesize$\gamma_3$};
\node at (gamma4) {\footnotesize$\gamma_4$};
\node at (gamma5) {\footnotesize$\gamma_5$};
\node at (gamma6) {\footnotesize$\gamma_6$};

\draw[->, >=latex] (gamma1) to (gamma5);
\draw[->, >=latex] (gamma1) to (gamma6);
\draw[->, >=latex] (gamma2) to (gamma5);
\draw[->, >=latex] (gamma2) to (gamma6);
\draw[->, >=latex] (gamma3) to (gamma1);
\draw[->, >=latex] (gamma3) to (gamma2);
\draw[->, >=latex] (gamma4) to (gamma1);
\draw[->, >=latex] (gamma4) to (gamma2);
\draw[->, >=latex] (gamma5) to (gamma3);
\draw[->, >=latex] (gamma5) to (gamma4);
\draw[->, >=latex] (gamma6) to (gamma3);
\draw[->, >=latex] (gamma6) to (gamma4);
\end{tikzpicture}& $576\color{lightgray}\times24\color{black}$
\\
\rule{0pt}{1.5cm}
$Q_3:$\begin{tikzpicture}[>=stealth, node distance=2cm, baseline={(current bounding box.center)}]

\node[circle, draw=black, inner sep=4pt] (gamma1) at (-0.5*0.75, 1.5*0.75) {};
\node[circle, draw=black, inner sep=4pt] (gamma2) at (0.5*0.75, 2*0.75) {};
\node[circle, draw=black, inner sep=4pt] (gamma3) at (0*0.75, 0*0.75) {};
\node[circle, draw=black, inner sep=4pt] (gamma4) at (2*0.75, 0*0.75) {};
\node[circle, draw=black, inner sep=4pt] (gamma5) at (1.5*0.75, 2*0.75) {};
\node[circle, draw=black, inner sep=4pt] (gamma6) at (2.5*0.75, 1.5*0.75) {};

\node at (gamma1) {\footnotesize$\gamma_4$};
\node at (gamma2) {\footnotesize$\gamma_2$};
\node at (gamma3) {\footnotesize$\gamma_5$};
\node at (gamma4) {\footnotesize$\gamma_6$};
\node at (gamma5) {\footnotesize$\gamma_3$};
\node at (gamma6) {\footnotesize$\gamma_1$};

\draw[->, >=latex] (gamma1) to (gamma4);
\draw[->, >=latex] (gamma2) to (gamma4);
\draw[->, >=latex] (gamma3) to (gamma1);
\draw[->, >=latex] (gamma3) to (gamma2);
\draw[->, >=latex] (gamma5) to (gamma3);
\draw[->, >=latex] (gamma6) to (gamma3);
\draw[->, >=latex] (gamma4) to (gamma5);
\draw[->, >=latex] (gamma4) to (gamma6);
\end{tikzpicture}& $1,656\color{lightgray}\times8\color{black}$
\\
\rule{0pt}{1.5cm}
$Q_4:$\begin{tikzpicture}[>=stealth, node distance=2cm, baseline={(current bounding box.center)}]

\node[circle, draw=black, inner sep=4pt] (gamma3) at (-0.75*0.75, -0.5*0.75) {};
\node[circle, draw=black, inner sep=4pt] (gamma4) at (1*0.75, -0.5*0.75) {};
\node[circle, draw=black, inner sep=4pt] (gamma2) at (2.5*0.75, -1.73*0.75) {};
\node[circle, draw=black, inner sep=4pt] (gamma6) at (1*0.75, -2.96*0.75) {};
\node[circle, draw=black, inner sep=4pt] (gamma5) at (2.75*0.75, -0.5*0.75) {};
\node[circle, draw=black, inner sep=4pt] (gamma1) at (-0.5*0.75, -1.73*0.75) {};

\node at (gamma3) {\footnotesize$\gamma_4$};
\node at (gamma4) {\footnotesize$\gamma_3$};
\node at (gamma2) {\footnotesize$\gamma_6$};
\node at (gamma6) {\footnotesize$\gamma_2$};
\node at (gamma5) {\footnotesize$\gamma_1$};
\node at (gamma1) {\footnotesize$\gamma_5$};

\draw[->, >=latex] (gamma1) to (gamma2);
\draw[->, >=latex] (gamma2) to (gamma3);
\draw[->, >=latex] (gamma2) to (gamma4);
\draw[->, >=latex] (gamma2) to (gamma5);
\draw[->, >=latex] (gamma6) to (gamma2);
\draw[->, >=latex] (gamma3) to (gamma1);
\draw[->, >=latex] (gamma4) to (gamma1);
\draw[->, >=latex] (gamma5) to (gamma1);
\draw[->, >=latex] (gamma1) to (gamma6);
\end{tikzpicture}\rule[-1.5cm]{0pt}{1.5cm}& $1,776\color{lightgray}\times6\color{black}$
\\
\hline
\end{tabular}
    \caption{The four BPS quivers of the $\text{SU(2)}$ $N_f =4$ SCFT and the number of minimal chambers identified by the RL model.}
    \label{fig:su2table}
\end{figure}
We note that for each finite chamber, there exists a set of equivalent chambers obtained from the symmetries of the quiver. One such symmetry is permutation of topologically equivalent nodes (formally an automorphic equivalence in graph theory). For $Q_1$ in Fig.~\ref{fig:su2table}, the nodes are arranged into classes
\begin{equation}
    \{\gamma_1\}\;,\{\gamma_2\}\;,\{\gamma_3,\gamma_4,\gamma_5,\gamma_6\}
\end{equation}
which introduces a $4!=24$-fold degeneracy to each chamber found for this quiver. The quivers $Q_2$ and $Q_3$ in Fig.~\ref{fig:su2table} also have global symmetries that introduce additional degeneracies, see \ref{app:symmetries}. In the counting procedure, the RL model does not identify these as distinct chambers, and so the results in Fig. \ref{fig:su2table} and Fig.~\ref{fig:all chambers}
should be interpreted as the number of essentially unique minimal chambers of each quiver.
\begin{figure*}[h]
\centering
\begin{tabular}{|c|c|}
\hline
BPS quiver & Minimal chamber \\
\hline
\rule{0pt}{1.5cm}
$\hat{\hat{E}}_6$:  \begin{tikzpicture}[>=stealth, node distance=2cm, baseline={(current bounding box.center)}]

\node[circle, draw=black, inner sep=4pt] (gamma1) at (0,0) {};
\node[circle, draw=black, inner sep=4pt] (gamma2) at (0,-1.5*0.75) {};
\node[circle, draw=black, inner sep=4pt] (gamma3) at (-1*0.75, -0.75*0.75) {};
\node[circle, draw=black, inner sep=4pt] (gamma4) at (1*0.75, 0) {};
\node[circle, draw=black, inner sep=4pt] (gamma5) at (1*0.75, -1.5*0.75) {};
\node[circle, draw=black, inner sep=4pt] (gamma6) at (-2*0.75, -0.75*0.75) {};
\node[circle, draw=black, inner sep=4pt] (gamma7) at (2*0.75, 0) {};
\node[circle, draw=black, inner sep=4pt] (gamma8) at (2*0.75, -1.5*0.75) {};

\node at (gamma3) {\footnotesize$\gamma_3$};
\node at (gamma4) {\footnotesize$\gamma_4$};
\node at (gamma2) {\footnotesize$\gamma_2$};
\node at (gamma6) {\footnotesize$\gamma_6$};
\node at (gamma5) {\footnotesize$\gamma_5$};
\node at (gamma1) {\footnotesize$\gamma_1$};
\node at (gamma7) {\footnotesize$\gamma_7$};
\node at (gamma8) {\footnotesize$\gamma_8$};

\draw[->, >=latex, transform canvas={xshift = -1pt}] (gamma1) to (gamma2);
\draw[->, >=latex, transform canvas={xshift=1pt}] (gamma1) to (gamma2);
\draw[->, >=latex] (gamma2) to (gamma5);
\draw[->, >=latex] (gamma2) to (gamma4);
\draw[->, >=latex] (gamma2) to (gamma3);
\draw[->, >=latex] (gamma3) to (gamma1);
\draw[->, >=latex] (gamma4) to (gamma1);
\draw[->, >=latex] (gamma5) to (gamma1);
\draw[->, >=latex] (gamma5) to (gamma8);
\draw[->, >=latex] (gamma6) to (gamma3);
\draw[->, >=latex] (gamma7) to (gamma4);
\end{tikzpicture}& 
$\begin{array}{c}
\text{12 states:}
\\
\gamma_8, \gamma_4, \gamma_3, \gamma_2, \gamma_1 + \gamma_3 + \gamma_4,\\
\gamma_2 + \gamma_5, \gamma_1 + \gamma_3, \gamma_6, \gamma_1 + \gamma_4, \gamma_7, \gamma_1, \gamma_5
\end{array}$
\\
\rule{0pt}{1.5cm}
$\hat{\hat{E}}_7$: \begin{tikzpicture}[>=stealth, node distance=2cm, baseline={(current bounding box.center)}]

\node[circle, draw=black, inner sep=4pt] (gamma1) at (0,0) {};
\node[circle, draw=black, inner sep=4pt] (gamma2) at (0,-1.5*0.75) {};
\node[circle, draw=black, inner sep=4pt] (gamma3) at (-1*0.75, -0.75*0.75) {};
\node[circle, draw=black, inner sep=4pt] (gamma4) at (1*0.75, 0) {};
\node[circle, draw=black, inner sep=4pt] (gamma5) at (1*0.75, -1.5*0.75) {};
\node[circle, draw=black, inner sep=4pt] (gamma7) at (2*0.75, 0) {};
\node[circle, draw=black, inner sep=4pt] (gamma8) at (2*0.75, -1.5*0.75) {};
\node[circle, draw=black, inner sep=4pt] (gamma72) at (3*0.75, 0) {};
\node[circle, draw=black, inner sep=4pt] (gamma9) at (3*0.75, -1.5*0.75) {};

\node at (gamma3) {\footnotesize$\gamma_3$};
\node at (gamma4) {\footnotesize$\gamma_4$};
\node at (gamma2) {\footnotesize$\gamma_2$};
\node at (gamma5) {\footnotesize$\gamma_5$};
\node at (gamma1) {\footnotesize$\gamma_1$};
\node at (gamma7) {\footnotesize$\gamma_6$};
\node at (gamma8) {\footnotesize$\gamma_8$};
\node at (gamma72) {\footnotesize$\gamma_7$};
\node at (gamma9) {\footnotesize$\gamma_9$};

\draw[->, >=latex, transform canvas={xshift = -1pt}] (gamma1) to (gamma2);
\draw[->, >=latex, transform canvas={xshift=1pt}] (gamma1) to (gamma2);
\draw[->, >=latex] (gamma2) to (gamma5);
\draw[->, >=latex] (gamma2) to (gamma4);
\draw[->, >=latex] (gamma2) to (gamma3);
\draw[->, >=latex] (gamma3) to (gamma1);
\draw[->, >=latex] (gamma4) to (gamma1);
\draw[->, >=latex] (gamma5) to (gamma1);
\draw[->, >=latex] (gamma5) to (gamma8);
\draw[->, >=latex] (gamma7) to (gamma4);
\draw[->, >=latex] (gamma72) to (gamma7);
\draw[->, >=latex] (gamma8) to (gamma9);
\end{tikzpicture} &
$\begin{array}{c}
\text{13 states:}\\
\gamma_4, \gamma_9, \gamma_8, \gamma_2, \gamma_2 + \gamma_5, \gamma_1 + \gamma_4, \gamma_1,\\ \gamma_6, \gamma_7, \gamma_5, \gamma_1 + \gamma_2 + \gamma_3, \gamma_2 + \gamma_3, \gamma_3
\end{array}$
\\
\rule{0pt}{1.5cm}
$\hat{\hat{E}}_8$: \begin{tikzpicture}[>=stealth, node distance=2cm, baseline={(current bounding box.center)}]

\node[circle, draw=black, inner sep=4pt] (gamma1) at (0,0) {};
\node[circle, draw=black, inner sep=4pt] (gamma2) at (0,-1.5*0.75) {};
\node[circle, draw=black, inner sep=4pt] (gamma3) at (-1*0.75, -0.75*0.75) {};
\node[circle, draw=black, inner sep=4pt] (gamma4) at (1*0.75, 0) {};
\node[circle, draw=black, inner sep=4pt] (gamma5) at (1*0.75, -1.5*0.75) {};
\node[circle, draw=black, inner sep=4pt] (gamma7) at (2*0.75, 0) {};
\node[circle, draw=black, inner sep=4pt] (gamma8) at (2*0.75, -1.5*0.75) {};
\node[circle, draw=black, inner sep=4pt] (gamma72) at (3*0.75, 0) {};
\node[circle, draw=black, inner sep=4pt] (gamma9) at (4*0.75, 0) {};
\node[circle, draw=black, inner sep=4pt] (gamma10) at (5*0.75, 0) {};

\node at (gamma3) {\footnotesize$\gamma_3$};
\node at (gamma4) {\footnotesize$\gamma_4$};
\node at (gamma2) {\footnotesize$\gamma_2$};
\node at (gamma5) {\footnotesize$\gamma_5$};
\node at (gamma1) {\footnotesize$\gamma_1$};
\node at (gamma7) {\footnotesize$\gamma_7$};
\node at (gamma8) {\footnotesize$\gamma_6$};
\node at (gamma72) {\footnotesize$\gamma_8$};
\node at (gamma9) {\footnotesize$\gamma_9$};
\node at (gamma10) {\footnotesize$\gamma_{10}$};

\draw[->, >=latex, transform canvas={xshift = -1pt}] (gamma1) to (gamma2);
\draw[->, >=latex, transform canvas={xshift=1pt}] (gamma1) to (gamma2);
\draw[->, >=latex] (gamma2) to (gamma5);
\draw[->, >=latex] (gamma2) to (gamma4);
\draw[->, >=latex] (gamma2) to (gamma3);
\draw[->, >=latex] (gamma3) to (gamma1);
\draw[->, >=latex] (gamma4) to (gamma1);
\draw[->, >=latex] (gamma5) to (gamma1);
\draw[->, >=latex] (gamma5) to (gamma8);
\draw[->, >=latex] (gamma7) to (gamma4);
\draw[->, >=latex] (gamma72) to (gamma7);
\draw[->, >=latex] (gamma9) to (gamma72);
\draw[->, >=latex] (gamma10) to (gamma9);
\end{tikzpicture} & 
$\begin{array}{c}
\text{14 states:}
\\
\gamma_4, \gamma_6, \gamma_2, \gamma_1 + \gamma_4, \gamma_7, \gamma_8, \gamma_9, \gamma_{10},\\ \gamma_2 + \gamma_3, \gamma_3, \gamma_1 + \gamma_3, \gamma_2 + \gamma_5, \gamma_1, \gamma_5
\end{array}$
\\
\rule{0pt}{1.5cm}
$X_6$: \begin{tikzpicture}[>=stealth, node distance=2cm, baseline={(current bounding box.center)}]

\node[circle, draw=black, inner sep=4pt] (gamma1) at (0,0) {};
\node[circle, draw=black, inner sep=4pt] (gamma2) at (1*0.75,1*0.75) {};
\node[circle, draw=black, inner sep=4pt] (gamma3) at (2*0.75, 0) {};
\node[circle, draw=black, inner sep=4pt] (gamma4) at (3*0.75, 1*0.75) {};
\node[circle, draw=black, inner sep=4pt] (gamma5) at (4*0.75, 0) {};
\node[circle, draw=black, inner sep=4pt] (gamma6) at (2*0.75, -1*0.75) {};

\node at (gamma3) {\footnotesize$\gamma_3$};
\node at (gamma4) {\footnotesize$\gamma_4$};
\node at (gamma2) {\footnotesize$\gamma_2$};
\node at (gamma6) {\footnotesize$\gamma_6$};
\node at (gamma5) {\footnotesize$\gamma_5$};
\node at (gamma1) {\footnotesize$\gamma_1$};

\draw[->, >=latex, transform canvas={yshift=1pt, xshift = -1pt}] (gamma1) to (gamma2);
\draw[->, >=latex, transform canvas={yshift=-1pt, xshift=1pt}] (gamma1) to (gamma2);
\draw[->, >=latex] (gamma3) to (gamma6);
\draw[->, >=latex] (gamma3) to (gamma1);
\draw[->, >=latex] (gamma3) to (gamma4);
\draw[->, >=latex] (gamma2) to (gamma3);
\draw[->, >=latex] (gamma3) to (gamma1);
\draw[->, >=latex, transform canvas={yshift=1pt, xshift = 1pt}] (gamma4) to (gamma5);
\draw[->, >=latex, transform canvas={yshift=-1pt, xshift = -1pt}] (gamma4) to (gamma5);
\draw[->, >=latex] (gamma5) to (gamma3);
\end{tikzpicture} &
$\begin{array}{c}
\text{10 states:}\\
\gamma_6, \gamma_2, \gamma_2 + \gamma_3, \gamma_5, \gamma_2 + \gamma_3 + \gamma_4,\\ \gamma_1, \gamma_2 + 2\gamma_3 + \gamma_4 + \gamma_5, \gamma_3 + \gamma_5, \gamma_4, \gamma_3
\end{array}$
\\
\rule{0pt}{1.5cm}
$X_7$: \begin{tikzpicture}[>=stealth, node distance=2cm, baseline={(current bounding box.center)}]

\node[circle, draw=black, inner sep=4pt] (gamma1) at (0,0) {};
\node[circle, draw=black, inner sep=4pt] (gamma2) at (1*0.75,1*0.75) {};
\node[circle, draw=black, inner sep=4pt] (gamma3) at (2*0.75, 0) {};
\node[circle, draw=black, inner sep=4pt] (gamma4) at (3*0.75, 1*0.75) {};
\node[circle, draw=black, inner sep=4pt] (gamma5) at (4*0.75, 0) {};
\node[circle, draw=black, inner sep=4pt] (gamma6) at (3*0.75, -1*0.75) {};
\node[circle, draw=black, inner sep=4pt] (gamma7) at (1*0.75, -1*0.75) {};

\node at (gamma3) {\footnotesize$\gamma_3$};
\node at (gamma4) {\footnotesize$\gamma_4$};
\node at (gamma2) {\footnotesize$\gamma_2$};
\node at (gamma6) {\footnotesize$\gamma_6$};
\node at (gamma5) {\footnotesize$\gamma_5$};
\node at (gamma1) {\footnotesize$\gamma_1$};
\node at (gamma7) {\footnotesize$\gamma_7$};

\draw[->, >=latex, transform canvas={yshift=1pt, xshift = -1pt}] (gamma1) to (gamma2);
\draw[->, >=latex, transform canvas={yshift=-1pt, xshift=1pt}] (gamma1) to (gamma2);
\draw[->, >=latex] (gamma3) to (gamma6);
\draw[->, >=latex] (gamma3) to (gamma1);
\draw[->, >=latex] (gamma3) to (gamma4);
\draw[->, >=latex] (gamma2) to (gamma3);
\draw[->, >=latex] (gamma3) to (gamma1);
\draw[->, >=latex, transform canvas={yshift=1pt, xshift = 1pt}] (gamma4) to (gamma5);
\draw[->, >=latex, transform canvas={yshift=-1pt, xshift = -1pt}] (gamma4) to (gamma5);
\draw[->, >=latex] (gamma5) to (gamma3);
\draw[->, >=latex, transform canvas={yshift=1pt}] (gamma6) to (gamma7);
\draw[->, >=latex, transform canvas={yshift=-1pt}] (gamma6) to (gamma7);
\draw[->, >=latex] (gamma7) to (gamma3);
\end{tikzpicture} \rule[-1.5cm]{0pt}{1.5cm}&
No finite chamber \\
\hline
\end{tabular}
\caption{The 5 exceptional complete $\mathcal{N}=2$ theories and their BPS quivers, and their minimal chamber as identified by the RL model. In agreement with the literature, we find that the $X_6$ Derksen-Owen quiver has no finite chamber.}
\label{fig: exceptional quivers}
\end{figure*}
For this experiment, the RL algorithm is adjusted to apply a negative penalty upon visiting a known chamber, which encourages the agent to further explore the moduli space and identify new minimal chambers. We run the searches until no new chambers are found in $\mathcal{O}(10^6)$ time-steps, which occurs after $\mathcal{O}(10^6\sim10^7)$ time-steps. The total identified for each quiver is shown in Fig. \ref{fig:su2table}, up to symmetries, and the full lists of these chambers are in the \href{https://github.com/fingriffin}{GitHub}. Each chamber can also be manually checked using the interactive quiver plot. The total number of minimal chambers found by the agent against total time-steps is shown in Fig.~\ref{fig:all chambers}.

\subsection{Exceptional complete theories}
The $\mathcal{N} =2$ models engineered from punctured Riemann surfaces \cite{Gaiotto:2009we} are all but finitely many of the complete theories with BPS quivers. We now divert our attention to identifying minimal chambers of the following exceptional complete theories, which are not constructed by the triangulation of a Riemann surface but rather by type IIB geometric engineering \cite{Alim:2011ae, Cecotti:2011rv}:

\begin{itemize}
\item The three theories identified with the elliptic $E$-type Dynkin diagrams, $\hat{\hat{E}}_n$ with $n = 6,7,8$ \cite{SaitoElliptic}. 
\item The two theories identified by the Derksen-Owen quivers, $X_6$ and $X_7$ \cite{DerksenOwen}.  
\end{itemize}
The RL algorithm identifies minimal chambers of these theories in $\mathcal{O}(10^2 \sim 10^3)$ timesteps, which are shown with their BPS quivers in Fig. \ref{fig: exceptional quivers}. In agreement with results from \cite{Alim:2011ae}, we find that the $X_7$ quiver has no finite chamber.
\section*{Acknowledgements}

FC would like to thank the Banff International Research Station and Cornell University for hospitality during the days in which this project was completed. AG would like to thank Alex Bloch for useful discussions on reinforcement learning. FC and YHH are supported by a Leverhulme Trust Research Project (Grant No. RPG-2022-145). FG was supported by the Crankstart Scholarship during the early stages of the project. 
\appendix
\section{Reinforcement Learning}
\label{RL}

\subsection{Proximal Policy Optimisation}
\label{PPO}
The dynamics of the intelligent agent are determined by a Markov decision process (MDP), formally a tuple \\ $(\mathcal{S},\mathcal{A},\{P(s,a)\},\gamma,R)$ where
\begin{itemize}
    \item $\mathcal{S}$ is the set of states in the environment.
    \item $\mathcal{A}$ is the set of actions permissible to the agent.
    \item $P(s,a)$ are the state transition probabilities. For each state $s\in\mathcal{S}$ and action $a\in\mathcal{A}$, this describes the distribution over what states the agent will transition to if the action $a$ is taken in state $s$.
    \item $\gamma \in  [0,1)$ is the discount factor.
    \item $R : \mathcal{S}\times\mathcal{A}\rightarrow \mathbb{R}$ is the reward function.
\end{itemize}
The MDP evolves as follows: starting with an initial state $s_0 \in \mathcal{S}$, the agent selects some action $a_0 \in \mathcal{A}$, drawn according to $P(s_0,a_0)$. This causes a transition from $s_0$ to some new state $s_1 \in \mathcal{S}$. The agent then picks another action $a_1$ drawn according to $P(s_1,a_1)$, which leads to $s_2 \in \mathcal{S}$. The process continues iteratively, with the agent choosing subsequent actions $a_0, a_1, a_2,\dots$, to navigate a trajectory of states:
\begin{equation}
s_0 \stackrel{a_0}{\longrightarrow} s_1 \stackrel{a_1}{\longrightarrow} s_2 \stackrel{a_2}{\longrightarrow} \dots
\end{equation}
The goal of any reinforcement learning algorithm is to find a policy $\pi: \mathcal{S}\rightarrow \mathcal{A}$ that maximises the so-called $Q$-function
\begin{equation}
Q^\pi(s_0) = E\bigl[R(s_0,a_0)+\gamma R(s_1,a_1)+\gamma^2 R(s_2,a_2) + \dots \;\big|\;\pi \bigr],
\end{equation}
which is the expected cumulative reward when the agent takes actions according to some policy $\pi$, i.e., $a_t = \pi(s_t)$.

Proximal Policy Optimisation (PPO) improves the policy $\pi$ by parameterising it with a neural network and introducing a clipped objective function to constrain policy updates. This stabilises training by ensuring gradual improvements. This robustness is particularly advantageous when combined with action masking to exclude domain-violating actions dynamically. Domain-violating actions are identified by the constraints of the environment, and we customise the model-free PPO algorithm to mask these actions dynamically. 

\subsection{$\boldsymbol{\epsilon}$-Greedy Strategy.} 
Although PPO generates a distribution over actions for each state, we further incorporate an $\epsilon$-greedy exploration scheme. In the language of the MDP tuple $(\mathcal{S}, \mathcal{A}, \{P(s,a)\}, \gamma, R)$, this approach modifies the effective policy distribution $\pi(s)$ from which actions are sampled. Concretely, with probability $\epsilon$, the agent chooses a random valid action (one that is not masked out by domain constraints), and with probability $1 - \epsilon$, it selects an action according to the PPO policy. This ensures sufficient exploration of the state-action space during training, mitigating premature convergence to suboptimal deterministic policies and helping the agent discover higher-value trajectories.

\subsection{Dynamic Action Masking}
\label{DAM}
Dynamic action masking is used to ensure that the agent never attempts actions outside the domain constraints defined by the environment. At each state $s$, a subset of actions in $\mathcal{A}$ may be prohibited due to physical laws, or human-imposed restrictions. Before the policy network computes the action probabilities for a given state $s$, a mask $\mathbf{m}(s)$ is generated, where $\mathbf{m}(s)$ is a binary vector of the same dimension as $\mathcal{A}$. Each entry of $\mathbf{m}(s)$ corresponds to an action in $\mathcal{A}$, taking the value $1$ if the action is valid and $0$ otherwise. This vector is dynamically constructed by querying the underlying environment constraints relevant to the current state. Formally, if $\boldsymbol{\ell}$ is the vector of logits for each action, the masked logits are given by
\[
\boldsymbol{\ell}' = \boldsymbol{\ell} \otimes \mathbf{m}(s),
\]
where $\otimes$ denotes element-wise multiplication. Logits associated with invalid actions are thus set to zero (or a sufficiently negative value if working in log space), ensuring that subsequent sampling of actions is limited to valid entries only.

\section{Graph automorphisms}
\label{app:symmetries}
An automorphism of a directed graph $G=(V,E)$ is a permutation $\sigma$ of the vertex set $V$ that preserves edge-vertex connectivity. That is, if the vertex pair $(u,v)$ is connected by a directed edge, the pair
\begin{equation}
(\sigma(u),\sigma(v))
\end{equation}
is also connected by an edge with the same direction. BPS quivers are directed graphs, with the vertex set $V$ the set of node labels $\{\gamma_i\}$. For each finite chamber of the quiver, there exists a set of equivalent chambers obtained through such automorphisms. The degeneracy of each chamber is therefore equal to the order of the automorphism group of the underlying quiver $\text{Aut}(G)$. The automorphism groups of the quivers from Fig. \ref{fig:su2table} are shown in Fig. \ref{fig: automorphisms} with the following considerations:
\begin{enumerate}
\item Pairs of nodes that are automorphically equivalent are identified with the same color. A quiver $G$ that contains $n$ such equivalent nodes contains a factor $S_n$ in $\text{Aut}(G)$.
\item Global symmetries of the graph $G$ represent automorphisms where entire classes of nodes are interchanged simultaneously. These contribute factors of $\mathbb{Z}_m$ to the group $\text{Aut}(G)$, with $m$ the order of the symmetry.
\end{enumerate}
The quivers $Q_1$ and $Q_4$ contain no such global symmetries. The quivers $Q_2$ and $Q_3$ have a third order rotational symmetry $\mathbb{Z}_3$, and a reflection symmetry $\mathbb{Z}_2$, respectively. Our code utilities the \texttt{networkx} package for Python to automatically detect these automorphisms for a given quiver.
\begin{figure}[h!]
\centering
\scalebox{0.9}{
\begin{tabular}{|c|c|c|}
\hline
BPS Quiver & $\text{Aut}(G)$ & Order \\
\hline
\rule{0pt}{1.5cm}
$Q_1:$ \begin{tikzpicture}[>=stealth, node distance=2cm, baseline={(current bounding box.center)}]

\node[circle, draw=black, inner sep=4pt, fill=cyan] (gamma3) at (-0.75*0.75, -0.5*0.75) {};
\node[circle, draw=black, inner sep=4pt, fill=cyan] (gamma4) at (2.75*0.75, -0.5*0.75) {};
\node[circle, draw=black, inner sep=4pt, fill=green] (gamma2) at (2.5*0.75, -1.73*0.75) {};
\node[circle, draw=black, inner sep=4pt, fill=cyan] (gamma6) at (2.75*0.75, -2.96*0.75) {};
\node[circle, draw=black, inner sep=4pt, fill=cyan] (gamma5) at (-0.75*0.75, -2.96*0.75) {};
\node[circle, draw=black, inner sep=4pt, fill=red] (gamma1) at (-0.5*0.75, -1.73*0.75) {};

\node at (gamma3) {\footnotesize$\gamma_3$};
\node at (gamma4) {\footnotesize$\gamma_4$};
\node at (gamma2) {\footnotesize$\gamma_2$};
\node at (gamma6) {\footnotesize$\gamma_6$};
\node at (gamma5) {\footnotesize$\gamma_5$};
\node at (gamma1) {\footnotesize$\gamma_1$};

\draw[->, >=latex, transform canvas={yshift=1pt}] (gamma1) to (gamma2);
\draw[->, >=latex, transform canvas={yshift=-1pt}] (gamma1) to (gamma2);
\draw[->, >=latex] (gamma2) to (gamma3);
\draw[->, >=latex] (gamma2) to (gamma4);
\draw[->, >=latex] (gamma2) to (gamma5);
\draw[->, >=latex] (gamma2) to (gamma6);
\draw[->, >=latex] (gamma3) to (gamma1);
\draw[->, >=latex] (gamma4) to (gamma1);
\draw[->, >=latex] (gamma5) to (gamma1);
\draw[->, >=latex] (gamma6) to (gamma1);
\end{tikzpicture}& $S_4$ & 24
\\
\rule{0pt}{1.5cm}
$Q_2:$\begin{tikzpicture}[>=stealth, node distance=2cm, baseline={(current bounding box.center)}]

\node[circle, draw=black, inner sep=4pt, fill=cyan] (gamma1) at (1.5*0.9+0.749*0.9, 0*0.9+1.06*0.9) {};
\node[circle, draw=black, inner sep=4pt, fill=cyan] (gamma2) at (1.5*0.9, 0*0.9+2*1.06*0.9) {};
\node[circle, draw=black, inner sep=4pt, fill=red] (gamma3) at (0*0.9,  0*0.9+2*1.06*0.9) {};
\node[circle, draw=black, inner sep=4pt, fill=red] (gamma4) at (0*0.9-0.749*0.9, 0*0.9+1.06*0.9) {};
\node[circle, draw=black, inner sep=4pt, fill=green] (gamma5) at (0*0.9, 0*0.9) {};
\node[circle, draw=black, inner sep=4pt, fill=green] (gamma6) at (1.5*0.9, 0*0.9) {};

\node at (gamma1) {\footnotesize$\gamma_1$};
\node at (gamma2) {\footnotesize$\gamma_2$};
\node at (gamma3) {\footnotesize$\gamma_3$};
\node at (gamma4) {\footnotesize$\gamma_4$};
\node at (gamma5) {\footnotesize$\gamma_5$};
\node at (gamma6) {\footnotesize$\gamma_6$};

\draw[->, >=latex] (gamma1) to (gamma5);
\draw[->, >=latex] (gamma1) to (gamma6);
\draw[->, >=latex] (gamma2) to (gamma5);
\draw[->, >=latex] (gamma2) to (gamma6);
\draw[->, >=latex] (gamma3) to (gamma1);
\draw[->, >=latex] (gamma3) to (gamma2);
\draw[->, >=latex] (gamma4) to (gamma1);
\draw[->, >=latex] (gamma4) to (gamma2);
\draw[->, >=latex] (gamma5) to (gamma3);
\draw[->, >=latex] (gamma5) to (gamma4);
\draw[->, >=latex] (gamma6) to (gamma3);
\draw[->, >=latex] (gamma6) to (gamma4);
\end{tikzpicture}& $S_2\times S_2\times S_2\times\mathbb{Z}_3$ & 24 
\\
\rule{0pt}{1.5cm}
$Q_3:$\begin{tikzpicture}[>=stealth, node distance=2cm, baseline={(current bounding box.center)}]

\node[circle, draw=black, inner sep=4pt, fill=cyan] (gamma1) at (-0.5*0.75, 1.5*0.75) {};
\node[circle, draw=black, inner sep=4pt, fill=cyan] (gamma2) at (0.5*0.75, 2*0.75) {};
\node[circle, draw=black, inner sep=4pt, fill=green] (gamma3) at (0*0.75, 0*0.75) {};
\node[circle, draw=black, inner sep=4pt, fill=lightgray] (gamma4) at (2*0.75, 0*0.75) {};
\node[circle, draw=black, inner sep=4pt, fill=red] (gamma5) at (1.5*0.75, 2*0.75) {};
\node[circle, draw=black, inner sep=4pt, fill=red] (gamma6) at (2.5*0.75, 1.5*0.75) {};

\node at (gamma1) {\footnotesize$\gamma_4$};
\node at (gamma2) {\footnotesize$\gamma_2$};
\node at (gamma3) {\footnotesize$\gamma_5$};
\node at (gamma4) {\footnotesize$\gamma_6$};
\node at (gamma5) {\footnotesize$\gamma_3$};
\node at (gamma6) {\footnotesize$\gamma_1$};

\draw[->, >=latex] (gamma1) to (gamma4);
\draw[->, >=latex] (gamma2) to (gamma4);
\draw[->, >=latex] (gamma3) to (gamma1);
\draw[->, >=latex] (gamma3) to (gamma2);
\draw[->, >=latex] (gamma5) to (gamma3);
\draw[->, >=latex] (gamma6) to (gamma3);
\draw[->, >=latex] (gamma4) to (gamma5);
\draw[->, >=latex] (gamma4) to (gamma6);
\end{tikzpicture}& $S_2\times S_2 \times\mathbb{Z}_2$ & 8
\\
\rule{0pt}{1.5cm}
$Q_4:$\begin{tikzpicture}[>=stealth, node distance=2cm, baseline={(current bounding box.center)}]

\node[circle, draw=black, inner sep=4pt, fill=cyan] (gamma3) at (-0.75*0.75, -0.5*0.75) {};
\node[circle, draw=black, inner sep=4pt, fill=cyan] (gamma4) at (1*0.75, -0.5*0.75) {};
\node[circle, draw=black, inner sep=4pt, fill=green] (gamma2) at (2.5*0.75, -1.73*0.75) {};
\node[circle, draw=black, inner sep=4pt, fill=lightgray] (gamma6) at (1*0.75, -2.96*0.75) {};
\node[circle, draw=black, inner sep=4pt, fill=cyan] (gamma5) at (2.75*0.75, -0.5*0.75) {};
\node[circle, draw=black, inner sep=4pt, fill=red] (gamma1) at (-0.5*0.75, -1.73*0.75) {};

\node at (gamma3) {\footnotesize$\gamma_4$};
\node at (gamma4) {\footnotesize$\gamma_3$};
\node at (gamma2) {\footnotesize$\gamma_6$};
\node at (gamma6) {\footnotesize$\gamma_2$};
\node at (gamma5) {\footnotesize$\gamma_1$};
\node at (gamma1) {\footnotesize$\gamma_5$};

\draw[->, >=latex] (gamma1) to (gamma2);
\draw[->, >=latex] (gamma2) to (gamma3);
\draw[->, >=latex] (gamma2) to (gamma4);
\draw[->, >=latex] (gamma2) to (gamma5);
\draw[->, >=latex] (gamma6) to (gamma2);
\draw[->, >=latex] (gamma3) to (gamma1);
\draw[->, >=latex] (gamma4) to (gamma1);
\draw[->, >=latex] (gamma5) to (gamma1);
\draw[->, >=latex] (gamma1) to (gamma6);
\end{tikzpicture}\rule[-1.5cm]{0pt}{1.5cm}& $S_3 $ & 6
\\
\hline
\end{tabular}
}
\caption{The four BPS quivers of the $\text{SU}(2)\;N_f=4$ SCFT and their automorphism groups.}
\label{fig: automorphisms}
\end{figure}
\newpage


\newpage
\bibliographystyle{elsarticle-num} 
\bibliography{bib}






\end{document}